\documentstyle[preprint,aps,eqsecnum]{revtex}
\begin{document}
\draft
\preprint{gr-qc/9402019  Alberta-Thy-55-93}

\title{Observables for spacetimes with two Killing
field symmetries}

\author{Viqar Husain}

\address{Theoretical Physics Institute,
University of Alberta\\ Edmonton, Alberta T6G 2J1, Canada.}

\maketitle

\begin{abstract}
\baselineskip=1em
  The Einstein equations for spacetimes with two commuting spacelike
Killing field symmetries are studied from a Hamiltonian point of
view.
The complexified Ashtekar canonical variables are used, and the
symmetry
reduction is performed directly in the Hamiltonian theory.  The
reduced
system corresponds to the field equations of the SL(2,R) chiral model
with additional constraints.

On the classical phase space, a method of obtaining an infinite
number
of constants of the motion, or observables, is given. The procedure
involves writing the Hamiltonian evolution equations as a
single `zero curvature' equation, and then employing  techniques
used in the study of two dimensional integrable models. Two infinite
sets  of  observables are obtained explicitly as functionals
of the phase space variables.  One set carries sl(2,R) Lie algebra
indices and forms an  infinite dimensional Poisson algebra,   while
the
other is formed from traces of SL(2,R) holonomies that commute with
one another. The restriction of the (complex) observables to the
Euclidean
and Lorentzian sectors is discussed.

It is also shown that the  sl(2,R) observables can be
associated with a solution generating technique which is linked to
that given by Geroch.

\end{abstract}

\vfill
\eject

\section{Introduction}

In classical general relativity one of the important questions
is that of finding exact solutions and extracting their
properties. This is hindered by the complexity of Einstein's
equations, and the discovery of a new solution is rare.

It is therefore usual to simplify the problem by seeking solutions
that have certain symmetries. These are normally specified
by requiring the metric to have a number of Killing vector fields,
which leads to a simplified set of equations to solve.

One such set of reduced equations is obtained by requiring the
metric to have two commuting  vector fields. This simplification
leads to a two dimensional field theory, and has the advantage that
it still leaves the gravitational field with two local degrees of
freedom, (unlike, for example the minisuperspace reductions, where
only a finite number of degrees of freedom remain). This symmetry
reduction
was first studied in detail by Geroch \cite{geroch}, who found that
the
resulting Einstein equations have an  infinite dimensional `hidden'
symmetry. These symmetry transformations of the equations provide a
solution generating technique, whereby, given one solution
with two commuting Killing fields, a new family of solution can be
generated. The solution generating technique was later presented
from other points of view \cite{kinnchit,hausern,wu}.
These equations have also been studied using the inverse scattering
 method \cite{belzak} to obtain solitonic solutions.

 The question of exact solutions is related to that of conserved
quantities. It is expected, as for any dynamical system, that exact
solutions will be labelled by values of the  conserved quantities.
In general relativity, for spacetimes with compact spacelike hyper
surfaces, the latter are also referred to as observables. This is
because if conserved quantities can be written explicitly as
functionals
of the phase space variables (which should always be possible),
they would also be the fully gauge invariant variables.

It is useful to have phase space observables for the classical
theory, in particular in attempts to prove integrability. For
example, in all the known two dimensional integrable models such as
the KdV and Sine-Gordon equations, an explicit generating procedure
for
observables may be used to prove integrability \cite{das,fadtakh}.

Apart from the classical questions, in attempts to construct a
canonical
quantum theory starting from general relativity, a complete set of
such
classical variables is a prerequisite
for certain quantization schemes, where  the quantum theory is to be
obtained as a representation of the  Poisson algebra of observables
\cite{ish,abhbook,kuch}. This  method has been under study for the
 nonperturbative approach to  quantum gravity using the Ashtekar
variables
 \cite{abhbook,ash} and the related loop space representation
\cite{lc}.
 It has been successful for the quantization of 2+1 gravity
\cite{2+1}.

 For the full Einstein equations, it is known
that the only `hidden' symmetries, apart from diffeomorphisms,
are constant rescalings of the metric \cite{tor1}. From this result
it follows that no observables can be built as integrals
of local functions of the initial data \cite{tor2}.
However,  from the works mentioned above, the two Killing field
reduced equations are known  to have an associated infinite
dimensional
symmetry group. It is then natural to ask what are the conserved
quantities associated with these symmetries, and in particular
what they are as functionals of the phase space variables.

In  some recent work \cite{nenadb}, a procedure based on methods used
for finding conservation laws for soliton
equations has been applied to the two Killing field reduced Einstein
equations. The starting point in this work was a particular form of
the metric with two commuting spacelike Killing fields. The dynamical
Einstein equations following from this were then studied using ideas
from
two dimensional integrable models.
If these quantities can be written as phase space functionals,
one would have an infinite number of  observables for this sector of
Einstein gravity. However it is not clear from this work how the
conserved quantities can be rewritten in terms of the ADM phase
space variables.

This paper addresses the question of obtaining observables for two
Killing
field reduced Einstein gravity. The main result presented below is
an explicit construction of an infinite number of phase space
observables for spacetimes with two commuting spacelike Killing
fields,
and with compact spatial hypersurfaces. The observables are
obtained for complexified gravity (i.e. complex phase space variables
  on a real manifold). The reality conditions
  are then discussed for the Euclidean and Lorentzian restrictions.

 The natural starting point is the Hamiltonian form of the
Einstein equations. The Ashtekar Hamiltonian formulation
\cite{abhbook,ash}
is used for this, and in the  next section the two Killing field
symmetry
is imposed in these variables to obtain a reduced first class
Hamiltonian
system which still has  two local degrees of freedom. This reduction
corresponds
to the Gowdy cosmological models \cite{gow}, and has been studied
earlier by the author and Smolin\cite{lv}. In the third section
the reduced system is fully gauge fixed, with the gauge fixing
conditions
chosen to put the Hamiltonian evolution equations in a suggestive
form.
This is discussed further in the following section, where a zero
curvature
form of the evolution equations is given.
The fifth section gives the procedure for obtaining the observables,
and
is based on methods used in two dimensional integrable models. There
is
also a discussion of the Poisson algebra of the observables.
The sixth section describes a solution generating technique for
this sector of the Einstein equations using the observables, and its
connection with the Geroch method.  The paper ends with a summary and
outlook for the quantization of this sector of gravity.

\section{Two Killing vector field reduction}

 The Ashtekar Hamiltonian variables for complexified general
relativity
 are the (complex) canonically conjugate pair $(A_a^i,
\tilde{E}^{ai})$
where $A_a^i$ is an so(3) connection and $\tilde{E}^{ai}$
is a densitized dreibein. $a,b,..$ are three dimensional spatial
indices
 and $i,j,..=1,2,3$ are internal so(3) indices.
The constraints of general relativity are
\begin{eqnarray}
{\cal G}^i &:=& D_a\tilde{E}^{ai} = 0,  \\
{\cal C}_a &:=& F_{ab}^i\tilde{E}^{ai}=0,  \\
{\cal H} &:=& \epsilon^{ijk}F_{ab}^i\tilde{E}^{aj}\tilde{E}^{bk}=0,
\end{eqnarray}
where $D_a\lambda^i = \partial_a \lambda^i +
\epsilon^{ijk}A_a^j\lambda^k$
is the covariant derivative, and $F_{ab}^i$ is its curvature.

Since the phase space variables are complex, reality conditions need
to be imposed to obtain the Euclidean or Lorentzian sectors. These
are
$A_a^i=\bar{A}_a^i$, $E^{ai}=\bar{E}^{ai}$ for the former,
and $A_a^i + \bar{A}_a^i = 2\Gamma_a^i(E)$,  $E^{ai}=\bar{E}^{ai}$
 for the latter. The $\Gamma_a^i(E)$ is the connection for spatial
 indices and the bar denotes complex conjugation.

 We now review the two commuting spacelike Killing field reduction of
these
constraints which was first presented in \cite{lv}. Working in
spatial
coordinates $x,y$, such that the Killing vector fields are
$(\partial/\partial x)^a$ and $(\partial/\partial y)^a$ implies that
the
phase space variables will depend on only one of the three spatial
coordinates. Specifically,  we assume that the spatial topology is
that of
a three torus so that the phase space variables depend on the time
coordinate  $t$ and one angular coordinate $\theta$. This situation
corresponds to one of the Gowdy cosmological models \cite{gow}.
 (The other permitted spatial topologies for the Gowdy cosmologies
 are $S^1\times S^2$ and  $S^3$.)

In addition to these Killing field conditions, we set to zero some
of the phase space variables as a part of the symmetry reduction:
\begin{eqnarray}
\tilde{E}^{x3}&=&\tilde{E}^{y3}=\tilde{E}^{\theta
1}=\tilde{E}^{\theta 2}=0,
 \nonumber \\
 A_x^3 &=& A_y^3=A_\theta^1 = A_\theta^2 = 0.
\end{eqnarray}
These conditions may be viewed as implementing a partial gauge fixing
and
solution to some of the constraints. The end result  below
(\ref{G}-\ref{H})
is  a simplified set of first class constraints which describes a two
dimensional field theory on $S^1\times R$ with two local degrees of
freedom.

Renaming the remaining variables
$A:=A_\theta^3$, $E:=\tilde{E}^{\theta 3}$ and  $A_\alpha^I$,
$\tilde{E}^{\alpha I}$, where $\alpha,\beta,..=x,y$ and $I,J,..=1,2$,
the reduced constraints are
\begin{eqnarray}
 G & := &\partial E + J =0, \label{G} \\
 C  & := & F_{\theta\alpha}E^{\alpha I} = 0,\label{C} \\
  H & := &-2\epsilon^{IJ}F_{\theta\alpha}^IE^{\alpha J}E
+ F_{\alpha\beta}E^{\alpha I} E^{\beta J}\epsilon_{IJ} \nonumber \\
  & = &-2 E E^{\alpha J} \epsilon^{IJ}\partial A^I_\alpha
 +2AEK - K_\alpha^\beta K_\beta^\alpha + K^2 = 0 \label{H},
\end{eqnarray}
where $\partial=(\partial/\partial\theta)$,
\begin{eqnarray}
K_\alpha^\beta:&=&A_\alpha^IE^{\beta I}, \ \ \ \ K:=K_\alpha^\alpha,
\\
J_\alpha^\beta:&=&\epsilon^{IJ} A_\alpha^I E^{\beta J} \ \ \ \
J:=J_\alpha^\alpha,
\end{eqnarray}
and $\epsilon^{12}=1=-\epsilon^{21}$.

The SO(3) Gauss law has been reduced to U(1) and the spatial
diffeomorphism
constraint to Diff($S^1$) as may be seen by calculating the Poisson
algebra
of the  constraints smeared by  functions $\Lambda$, $V$, and the
lapse $N$
(which is a density of weight -1):
\begin{eqnarray}
G(\Lambda) &=& \int_0^{2\pi} d\theta\ \Lambda G, \\
C(V) &=& \int_0^{2\pi} d\theta\ VC, \\
H(N) &=&\int_0^{2\pi} d\theta\ NH
 \end{eqnarray}
\begin{eqnarray}
\{G(\Lambda),G(\Lambda')\} &=& = \{G(\Lambda),H(N)\}=0,\\
\{C(V),C(V')\}&=&C({\cal L}_VV' ), \\
\{H(N),H(N')\}&=& C(W) - G(AW) ,
\end{eqnarray}
where
\begin{equation}
W\equiv E^2(N\partial N' - N'\partial N).  \label{stfn}
\end{equation}
This shows that $C$ generates Diff$(S^1)$. Also we note that this
reduced
system still describes a sector of general relativity due to the
Poisson
bracket $\{H(N),H(N')\}$,  which is the reduced version of that for
full
general relativity  in the Ashtekar variables.

The variables $K_\alpha^\beta$ and $J_\alpha^\beta$ defined above
will be used below in the discussion of observables. Here we note
their
properties. They are invariant under the reduced Gauss law (\ref{G}),
transform as densities of weight +1 under the Diff($S^1$) generated
by $C$,
and form the   Poisson algebra
\begin{eqnarray}
\{K_\alpha^\beta,K_\gamma^\sigma\} &=& \delta_\alpha^\sigma
K_\gamma^\beta
 - \delta_\gamma^\beta K_\alpha^\sigma, \\
\{J_\alpha^\beta,J_\gamma^\sigma\} &=& -\delta_\alpha^\sigma
K_\gamma^\beta
 + \delta_\gamma^\beta K_\alpha^\sigma, \\
 \{K_\alpha^\beta,J_\gamma^\sigma\} &=& \delta_\alpha^\sigma
J_\gamma^\beta
 - \delta_\gamma^\beta J_\alpha^\sigma.
\end{eqnarray}
This shows that  $K_\alpha^\beta$ form the gl(2) Lie algebra, and
hence
generate gl(2) rotations on variables with indices
$\alpha,\beta,..=x,y$.

The following linear combinations of $K_\alpha^\beta$ form the
sl(2,R)
subalgebra of gl(2,R):
 \begin{equation} L_1 = {1\over 2}(K_y^x + K_x^y) \ \
  L_2 = {1\over 2}(K_x^x - K_y^y) \ \
  L_3 = {1\over 2}(K_y^x - K_x^y)
  \end{equation}
The Poisson bracket algebra of these is
\begin{equation} \{L_i,L_j\} = C_{ij}^{\ \ k}L_k,
 \end{equation}
 where $C_{12}^{\ \ 3} = -1, C_{23}^{\ \ 1} = 1, C_{31}^{\ \ 2} = 1$
are
 the sl(2,R) structure constants.
The corresponding linear combinations  of $J_\alpha^\beta$ are
denoted
 by $J_i$, i=1,2,3. Their Poisson brackets  are
 \begin{equation}
 \{L_i,J_j\}=C_{ij}^{\ \ k}J_k,\ \ \
 \{J_i,J_j\}= - C_{ij}^{\ \ k}L_k.
 \end{equation}
 Also
 \begin{equation}
 \{J,J_i\}=\{J,L_i\}=\{K,J_i\}=\{K,L_i\}= 0.
 \end{equation}
For discussing observables, it will turn out to be very convenient to
replace the eight canonical phase space variables
$A_\alpha^I,\tilde{E}^{\alpha I}$ by the eight Gauss law invariant
variables
$K_\alpha^\beta, J_\alpha^\beta$.

\section{Gauge fixing and the metric}

The Dirac observables are defined as the phase space functionals
$O[A,E]$ that have vanishing Poisson brackets with all the first
class constraints of the theory. This is because the first class
constraints generate local gauge transformations via
Poisson brackets. The question of finding the observables can be
equally well addressed prior to, or after, full gauge fixing of a
first
class system. Each will yield observables in terms of the phase space
variables.

Assuming that variables $O[A,E]$ invariant under the kinematical
Gauss
law and spatial diffeomorphism invariant  have already been
determined,
(which is relatively easy), the first case would
correspond to solving  for $O[A,E]$ the equation:
\begin{equation}
\{H(N),O\}\sim 0.
\end{equation}
 The second amounts to solving
\begin{equation}
\{\tilde{H},O\} = 0
\end{equation}
where the last equality is strong, and $\tilde{H}$ is a suitably
gauge
fixed Hamiltonian constraint.
The second procedure will be followed
here since, with a particular gauge choice to be described in this
section,
the Hamiltonian evolution equations can be put in a very simple form.

Full gauge fixing using the Ashtekar variables requires a careful
consideration of the reality conditions on the phase space coordinate
$A_a^i$. This is because the (complex) phase space variables depend
on
 real coordinates. For conventional gauge fixing where some functions
of the phase space variables are chosen as the coordinates, real
functions must be chosen.  But since the constraints themselves
are complex, two real conditions must be imposed for complete gauge
fixing. Here we gauge fix the {\it complex} theory by requiring that
certain (complex) functions of the phase space variables vanish. This
results in  (complex) gauge fixed evolution equations and
second class constraints. The reality conditions are discussed below,
where the metric resulting from the gauge fixing is compared with
the standard metric for this reduction, and in section V where the
observables are obtained.

 We start by fixing the Gauss law (\ref{G}) by imposing the gauge
fixing
 condition $A=0$. Solving this constraint gives
 \begin{equation}
  E = c - \int^\theta d\theta' J(\theta'),
 \end{equation}
 where $c$ is an arbitrary constant.
 The  diffeomorphism and Hamiltonian constraints (\ref{C}- \ref{H})
in
this gauge become
 \begin{eqnarray}
   H &=& -2 (c-\int^\theta d\theta'\ J(\theta')) E^{\alpha J}
   \epsilon^{IJ}\partial A^I_\alpha
  - K_\alpha^\beta K_\beta^\alpha + K^2 \label{H2} \\
  C &=& E^{\alpha I}\partial A_\alpha^I. \label{C2}
\end{eqnarray}
 and are still first class. In particular (\ref{H2}) satisfies the
 Poisson bracket relation
 \begin{equation}
 \{H(N),H(N')\} =  C(W),
 \end{equation}
 with $W$ given by (\ref{stfn}), which is the usual Poisson
 bracket of the Hamiltonian constraint with itself.
 Thus (\ref{H2}-\ref{C2}) on the $A_\alpha^I,\tilde{E}^{\alpha I}$
 phase space still describe general relativity with two
 local degrees of freedom.

We now work with the eight Gauss law invariant densities
$L_i, J_i$ and $K,J$ introduced in the last section
instead of the eight remaining phase space variables
$A_\alpha^I,\tilde{E}^{\alpha I}$. The evolution equations
$\dot{F} = \{F,H(N)\}$ with $H$ from (\ref{H2}) for these variables
are
\begin{eqnarray}
\dot{L}_i &=& -2\partial
\bigl[ N(c-\int^\theta d\theta'\ J(\theta'))J_i \bigr],
\label{evo1} \\
\dot{J}_i &=& 2\partial
\bigl[ N(c-\int^\theta d\theta'\ J(\theta'))L_i \bigr]
  +  4N C_i^{\ jk}J_jL_k, \label{evo2}
\end{eqnarray}
and
\begin{eqnarray}
\dot{J} &=& 2 \partial
\bigl[ N(c-\int^\theta d\theta'\  J(\theta'))K \bigr],
\label{evo3}\\
\dot{K} &=& -2 \partial
\bigl[ N(c-\int^\theta d\theta'\  J(\theta'))J \bigr].
\label{evo4}
\end{eqnarray}

A natural and consistent gauge fixing for the remaining gauge freedom
is achieved by the choice $J=0,K=constant$, so that the
diffeomorphism and
Hamiltonian constraints become strongly zero.

The condition $J=0$ is second class
with the Hamiltonian constraint, and in this gauge the evolution
equations (\ref{evo3}-\ref{evo4}) for $J,K$  reduce to
\begin{equation}
\dot{K}=0, \ \ \ \ \ \ \ \ \dot{J}=0=2c\partial(NK).
\end{equation}
 The first implies that $K=K(\theta)$. The second fixes the
lapse to be $N(\theta)=a/K(\theta)$, where $a$ is a complex constant.
We may now fix the (real) $\theta$ coordinate condition by setting
$ReK(\theta)=k\ne 0$ and $ImK(\theta)=0$, where $k$ is a real
constant
density on the circle.  Thus these gauge fixing
conditions on the phase space variables imply that the lapse and
shift
functions are constants. If we choose $a$ to be real, the lapse is
real.

The evolution equations (\ref{evo1}-\ref{evo2}) for the remaining
variables,  the six $L_i,J_i$,  with the   choice $N=1/4,c=2$ become
 \begin{equation}
 \dot{L}_i + J_i ' = 0
 \label{con}\end{equation}
 \begin{equation}
 \dot{J}_i - L_i'  +   C_i^{\ jk}L_jJ_k = 0.
 \label{cur}
 \end{equation}
where $'\equiv \partial /\partial \theta $. These, together with the
strongly imposed Hamiltonian and diffeomorphism constraints
 \begin{eqnarray}
  \epsilon^{IJ} E^{\alpha J} \partial A^I_\alpha
  + L_1^2 + L_2^2 - L_3^2 &=& 0
 \label{sc1} \\
   E^{\alpha I}\partial A_\alpha^I &=& 0, \label{sc2}
 \end{eqnarray}
form the fully gauge fixed set of two Killing field reduced
complex Einstein equations. There are  6-2 = 4 local phase space
degrees
of freedom. We note that these are written in terms of the original
phase space variables, so that Poisson brackets may still be
calculated using the fundamental $(A_\alpha^I,E^{\alpha I})$
bracket. Also, the gauge fixing has reduced the gl(2,R) Casimir term
 in the Hamiltonian constraint to the sl(2,R) Casimir (\ref{sc1}).

 We have  used the gauge conditon $J=0$ which is not
explicitly time dependent and is second class with the Hamiltonian
constraint. Normally such a gauge condition
for compact spatial hypersurfaces implies that the lapse function
must
be zero, which implies no evolution and a degenerate metric.
But this is not the case for the two Killing field reduction with
this gauge
due to (\ref{evo3}-\ref{evo4}), which are consistent with constant
non-zero
lapse, and as we see below, lead to a metric of the standard type for
two Killing field reductions. The two conditions $J=0,K=constant$
imply
 that the shift is also constant.

The main purpose of the gauge fixing was not to get an explicitly
reduced Hamiltonian in terms  of the two physical degrees of freedom,
but  to look at the full evolution equations (\ref{evo1}-\ref{evo2})
in a  convenient gauge which is useful for obtaining  the conserved
quantities.  The $J=0$ gauge is very convenient for this.
However one can obtain a non-vanishing reduced  Hamiltonian as a
function
of $L_i,J_i$ that leads to the evolution equations
(\ref{con}-\ref{cur}).
It is that for the Sl(2,R) chiral model.

Since the gauge fixed evolution equations (\ref{con}-\ref{cur})
involve only $J_i,L_i$, the conserved charges will depend only
on these. It is therefore important to check that the charges commute
with the second class constraints (\ref{sc1}-\ref{sc2}).
The commutation with the strong Hamiltonian constraint is guaranteed
because the variables $J_i,L_i$ commute with $J,K$, (which are the
variables fixed in the gauge choice):
 \begin{equation}
 \dot{L}_i = \{ L_i, H(N) \}|_{J=0\atop K=const.} =
  \{  L_i, H(N)|_{J=0\atop K=const.} \},
 \end{equation}
with the same equation holding for $J_i$. This is another reason
why it appears natural to  separate  the phase space variables into
sl(2,R)
variables $J_i,L_i$, with gauge conditions imposed on the traces
$J,K$.
As we will see below, the charges also commute with the
diffeomorphism
constraint (\ref{sc2}) by construction.

We will not solve the second class constraints explicitly, since the
goal is only to obtain the conserved quantities. The second class
constraints imply that there are two relations among the six
$J_i,L_i$.
In principle these can be substituted into the conserved quantities
to
rewrite them in terms of four independent reduced variables.

For comparison with the usual metric variables, it is useful to see
what form of the metric arises from the gauge choices made above.
The  general line element with two commuting spacelike Killing fields
can be put in the form \cite{gow}
\begin{equation}
ds^2 = e^{2F}(-dt^2 + d\theta^2) + g_{\alpha\beta}dx^\alpha dx^\beta
\label{smet}
\end{equation}
 where the four functions
$F,g_{\alpha\beta}$ are four functions of $t,\theta$ only.
On the other hand, the gauge choices made above lead to the line
element
\begin{equation}
ds^2 = -({1\over 16} + {1\over 2\sqrt{q}} C^2)dt^2 +
 {C\over \sqrt{q}}dt d\theta + {1\over 2\sqrt{q}} d\theta^2 +
{2\over \sqrt{q}} q_{\alpha\beta}dx^\alpha dx^\beta
\label{gmet}
\end{equation}
where $q_{\alpha\beta}$ is the matrix inverse of $\tilde{E}^{\alpha
I}
\tilde{E}^{\beta I}$,
 $q = det q_{\alpha\beta}$, and $C$ is a constant (the shift).
By a suitable gauge condition which fixes
$F$ as a function of the other three metric variables
$g_{\alpha\beta}$,
and a coordinate transformation, (\ref{smet}) can be brought into the
form (\ref{gmet}). In arriving at (\ref{gmet}), $E^{ai}$ has been
fixed
to be real (reality condition),  and the lapse and shift chosen
to be real constants. Note that while the reality conditions on
$A_a^i$ have not been imposed, this does not affect the general form
(\ref{gmet}) of the Lorentzian metric that will result.

We now note an alternative natural gauge fixing which may also be
useful
for this system but will not be used in this paper. The Hamiltonian
constraint (\ref{H}) contains the product $AE$, and $E$ transforms
like
 a scalar under the reduced diffeomorphism constraint (\ref{C}).
 This suggests the (partial) gauge fixing $ReE=t$, $ImE=0$,
  which gives  $H_R:=-A$ as the (complex) reduced
  Hamiltonian. Substituting this gauge condition into
(\ref{G}-\ref{H})
gives the first class constraints
\begin{eqnarray}
J &=& 0, \\
E^{\alpha I}\partial A_\alpha^I &=& 0,
\end{eqnarray}
and the time dependent reduced Hamiltonian
\begin{equation}
H_R = -{1\over K} E^{\alpha J} \epsilon^{IJ}\partial A^I_\alpha
  + {1\over 2Kt}\bigl( K^2 - K_\alpha^\beta K_\beta^\alpha \bigr).
\end{equation}
The time dependence in $H_R$ is associated only with the ultralocal
part,
 which is also the gl(2,R) Casimir invariant. This suggests that for
 small times the ultralocal piece dominates the dynamics and that
 a perturbation theory in $t$ may be possible.  The reality
 conditions on the $A$'s still need to be applied.

\section{Evolution equations as a zero curvature condition}

The evolution equations (\ref{con}-\ref{cur}) derived in the last
section
can be rewritten in a compact form using the sl(2,R) matrix
generators
\begin{equation} g_1 = {1\over 2}\left( \begin{array}{cc}
         0 & 1 \\
         1 & 0 \end{array} \right) \ \ \
   g_2 = {1\over 2}\left( \begin{array}{cc}
         1 & 0 \\
         0 & -1 \end{array} \right)\ \ \
   g_3 = {1\over 2}\left( \begin{array}{cc}
         0 & 1 \\
	 -1 & 0 \end{array} \right)
\end{equation}
which satisfy  the relations
$[g_i,g_j]=C_{ij}^{\ \ k}g_k$ and $g_ig_j={1\over 2}C_{ij}^{\ \
k}g_k$.
Defining the matrices
\begin{equation}
 A_0 := L_ig_i \ \ \ \ \ \ A_1 := J_ig_i,
 \end{equation}
  the evolution equations (\ref{con}-\ref{cur}) become
\begin{equation} \partial_0 A_0 + \partial_1 A_1 = 0
 \label{cons}
 \end{equation}
\begin{equation} \partial_0 A_1 - \partial_1 A_0 + [A_0, A_1] = 0.
\label{curv}\end{equation}
Equations (\ref{cons}-\ref{curv}) are the first order form of the
SL(2,R) chiral model field equations.

The two evolution equations (\ref{cons}-\ref{curv}) may be rewritten
as a single equation in the following way. Define for a real
parameter
$\lambda$
\begin{equation}
a_0:= {1\over 1+\lambda^2} \bigl( A_0 - \lambda A_1 \bigr) \ \ \
a_1:= {1\over 1+\lambda^2} \bigl( \lambda A_0 +  A_1 \bigr)
\end{equation}
Then equations (\ref{cons}-\ref{curv}) follow from the
single `zero curvature' equation
\begin{equation}
\dot{a_1} - a_0' + [a_0,a_1] = 0.
 \label{zcurv}
\end{equation}
This equation, together with the strong constraints
(\ref{sc1}-\ref{sc2})
form the two spacelike commuting Killing field reduction. The
dynamical
equation (\ref{zcurv}) is used in the following section to obtain the
conserved charges.

\section{Observables}

The field equations for all the known two dimensional integrable
models
have  zero curvature formulations analagous to that given in the last
section.
This is a direct consequence of the existence of two distinct
symplectic
forms on the phase spaces of the models \cite{das}, which is also the
geometric way of viewing the Lax pair formulation. Another
consequence of the
zero curvature formulation is a  procedure for generating an infinite
number
of conserved charges. We now apply this to the dynamical equation
(\ref{zcurv}) arising from the two Killing field reduction.
The resulting observables will be for complex gravity and the reality
conditions on them will be discussed at the end of the section.

 The transfer matrix used in the study of two dimensional models is
 analagous to the Wilson loop. For the present case, it is the path
 ordered exponential associated with the matrix $a_1$:
 \begin{equation}
 U[A_0,A_1](0,\theta) :=
\lim_{N\rightarrow\infty\atop\Delta\theta\rightarrow
0}
  \prod_{i=0}^N [1+a_1(\theta_i)\Delta\theta] \equiv
 {\rm Pexp}\int_0^{\theta} a_1(A_0,A_1,\lambda)\ d\theta '
 \end{equation}
The trace of the transfer matrix is preserved under time
evolution as may be seen by  deriving its equation of motion using
equation (\ref{zcurv}). We note first that
\begin{equation}
U'(0,\theta) = U(0,\theta)a_1(\theta),\ \ \ \ \
U'(\theta,2\pi) = -a_1(\theta)U(\theta,2\pi).
\end{equation}
 The time derivative of the first gives
\begin{eqnarray}
\dot{U}'(0,\theta) &=& \dot{U}(0,\theta)a_1 + U(0,\theta)\dot{a}_1
\nonumber \\
&=& \dot{U}(0,\theta)a_1 +  U(0,\theta)(a_0' - [a_0,a_1]),\nonumber
\end{eqnarray}
which may be rewritten as
\begin{equation}
(\dot{U}(0,\theta) - U(0,\theta)a_0)'
= (\dot{U}(0,\theta) - U(0,\theta)a_0)a_1.
\end{equation}
Thus, since  $\dot{U}(0,\theta) - U(0,\theta)a_0$ satisfies the same
equation as $U(0,\theta)$, we get the equation of motion
\begin{equation}
\dot{U}(0,\theta) = U(0,\theta)a_0(\theta) -  a_0(0) U(0,\theta).
\end{equation}
 From this it follows that
\begin{equation}
M[A_0,A_1](\lambda) := {\rm Tr}U(0,2\pi)
\label{mon}
\end{equation}
is conserved in time. The conservation of this trace  follows in
basically the same way as the conservation of the  Wilson loop
observable
when there is a zero-curvature constraint on the phase space, such as
in 2+1 gravity \cite{2+1}. That $M$ is spatial diffeomorphism
invariant
follows from noting that $a_1$ transforms like a density under the
Diff$(S^1)$ generated by (\ref{C2}):
\begin{equation}
\{ C(V), a_1 \} = -\partial(Va_1), \nonumber
\end{equation}
from which it follows that $\{ C(V), M \} = 0 $.

Expanding  $M$ in a power series in $\lambda$  gives explicitly
the phase space observables, which are the coefficients of
powers of $\lambda$:
The first three observables are
\begin{equation}
Q^0:= M|_{\lambda = 0} = {\rm TrPexp}[\int_0^{2\pi} d\theta\ A_1]
=: {\rm Tr}V(0,2\pi),
\label{lam0}\end{equation}
\begin{equation}
Q^1:= {\partial M\over \partial \lambda}|_{\lambda = 0} =
\int_0^{2\pi} d\theta\  {\rm
Tr}[V(0,\theta)A_0(\theta)V(\theta,2\pi)],
 \label{lam1}
 \end{equation}
 and
\begin{eqnarray}
{\partial^2 M\over \partial \lambda^2}|_{\lambda = 0} &=&
-2 \int_0^{2\pi} d\theta\   {\rm
Tr}[V(0,\theta)A_1(\theta)V(\theta,2\pi)]
\nonumber \\
& & + \int_0^{2\pi} d\theta_1\  \int_0^{2\pi} d\theta_2\
\Theta(\theta_2-\theta_1) \nonumber \\
& & {\rm Tr}[V(0,\theta_1)A_0(\theta_1)V(\theta_1,\theta_2)
A_0(\theta_2) V(\theta_2,2\pi)],
\label{lam2}
\end{eqnarray}
 where $\Theta(\theta-\theta') = 1, \theta\geq\theta'$ and zero
otherwise.
 It is straightforward to verify directly the conservation of these
 functionals using the equations of motion (\ref{cons}-\ref{curv}).

 The structure of the general observable can now be seen and we can
write
 down the observable with $n$ insertions of $A_0$ in the holonomies
  $V$:
  \begin{eqnarray}
Q^{n} &:=& \int_0^{2\pi} d\theta_1...\int_0^{2\pi} d\theta_n
\Theta(\theta_n-\theta_{n-1})...\Theta(\theta_2-\theta_1) \nonumber
\\
& &{\rm
Tr}[V(0,\theta_1)A_0(\theta_1)V(\theta_1,\theta_2)A_0(\theta_2)...
A_0(\theta_n)V(\theta_n,2\pi)].
 \label{genob}
  \end{eqnarray}
 This has a remarkable resemblance to the $T$  variables used in
 3+1 gravity \cite{lc},
 \begin{equation}
 T^{a_1...a_n}[A_a^i, \tilde{E}^{ai}](x_1,...x_n;\alpha):=
 {\rm Tr}[U_\alpha(x_0,x_1)\tilde{E}^{a_1}(x_1)U_\alpha(x_1,x_2)...
 \tilde{E}^{a_n}(x_n)U_\alpha(x_n,x_0)],
 \end{equation}
 where the holonomies $U_\alpha$ are based on the loop $\alpha$ are
made
 from
 Ashtekar's  connection $A_a^i$, and the insertions in the product of
 holonomies are the conjugate momenta $\tilde{E}^{ai}$ instead of
$A_0$.
 The other difference is that in equation (\ref{genob}) there is an
integration over all the point insertions of $A_0$, (which gives
invariance under the remaining spatial diffeomorphisms Diff$(S^1)$ in
the present reduction).

Another set of observables is obtained  by looking at the first
 term in (\ref{lam2}) where there is an insertion of $A_1$ in the
holonomies instead of $A_0$. The general observables  of this type
is similar to (\ref{genob})  but with  $n$ insertions
 of $A_1$:
  \begin{eqnarray}
 P^{n} &:=& \int_0^{2\pi}d\theta_1...\int_0^{2\pi}d\theta_n
 \Theta(\theta_n-\theta_{n-1})...\Theta(\theta_2-\theta_1) \nonumber
\\
   & &
Tr[V(0,\theta_1)A_1(\theta_1)V(\theta_1,\theta_2)A_1(\theta_2)...
    A_1(\theta_n)V(\theta_n,2\pi)].
\label{genob2}
\end{eqnarray}

 The Poisson algebra of the observables (\ref{genob}-\ref{genob2})
 may be calculated using the trace identity
 \begin{equation}
{\rm Tr}[Xg^i]{\rm Tr}[Yg^j]=C_k^{\ ij}
({\rm Tr}[Xg^kY]-{\rm Tr}[Yg^kX]),
\nonumber
\end{equation}
for SL(2,R) matrices $X,Y$ and generators $g^i$. We find
\begin{eqnarray}
\{ P^m,P^n\}&=&0 \\
\{ Q^m,Q^n\} &\sim& Q^{m+n-1} + Q^{m+n-1},\ m,n>1 \\
\{ Q^0,Q^m\}&\sim& Q^{m+1}
\end{eqnarray}

 There is another  method for generating  conserved charges for
two dimensional chiral models \cite{bizz} which can be applied here
to
generate observables. This is useful for comparison with the above
procedure. Also,  as discussed below, the resulting observables for
the Killing field reduction  give a
solution generating technique which may be viewed as the
Hamiltonian analog of Geroch method \cite{geroch}. This procedure for
obtaining observables has also
been applied to self-dual gravity \cite{sdvh,sdstr}.

The starting point is the dynamical equations
(\ref{cons}-\ref{curv}). We note that  (\ref{cons}) is already
 like a conservation law and so the first conserved charge is
 \begin{equation}
 q^{(1)} = \int_0^{2\pi} d\theta\ A_0 = \int_0^{2\pi} d\theta\
L_ig_i,
 \end{equation}
 which gives the three sl(2,R) charges
 \begin{equation}
 q^{(1)}_i =  \int_0^{2\pi} d\theta\ L_i.
 \end{equation}
 These observables were obtained earlier in \cite{lv}.

The current ${\cal J}_\mu^{(1)} := A_\mu$ ($\mu,\nu,..=0,1)$ is
conserved
so there exists a (matrix) function  $f^{(1)}(t,\theta)$ such that
\begin{equation}
{\cal J}_\mu^{(1)} = \epsilon_\mu^{\ \nu}\partial_\nu f^{(1)}.
\end{equation}
 We  now define  the second current  by
\begin{equation}
{\cal J}_\mu^{(2)} := D_\mu  f^{(1)} \equiv \partial_\mu f^{(1)} +
A_\mu f^{(1)}
\end{equation}
With this definition of a derivative operator, the equation of motion
 (\ref{curv}) may be rewritten as $[D_0,D_1]=0$.
 The  conservation of  ${\cal J}_\mu^{(2)}$ is easy to show:
\begin{eqnarray}
\delta^{\mu\nu} \partial_\nu {\cal J}_\mu^{(2)} &=& \delta^{\mu\nu}
\partial_\nu D_\mu  f^{(1)}  = \delta^{\mu\nu}D_\mu\partial_\nu
f^{(1)}
\nonumber \\
& = & \delta^{\mu\nu}\epsilon^{\mu\alpha}D_\mu {\cal J}_\alpha^{(1)}
 = \delta^{\mu\nu}\epsilon^{\mu\alpha}D_\mu D_\alpha f^{(0)} = 0,
\end{eqnarray}
where the last  equality follows because
${\cal J}_\mu^{(1)}= D_\mu f^{(0)} = A_\mu, $
where $f^{(0)}$ is the identity matrix, and $[D_0,D_1]=0$ by the
equation
of motion (\ref{curv}).
This procedure generalizes, and it is straight forward to give an
inductive proof that
${\cal J}^{(n+1)} := D_\mu f^{(n)}$ is conserved, assuming
${\cal J}^{(n)}_\mu$ is conserved. The observables are
\begin{equation}
q^{(n)} := \int_0^{2\pi} d\theta\ {\cal J}^{(n)}_0.
\end{equation}

 The second conserved charge is
\begin{equation}
q^{(2)} := \int_0^{2\pi} d\theta\ D_0  f^{(1)}(\theta,t)
= \int_0^{2\pi} d\theta\ \bigl( - A_1(\theta,t) + A_0 (\theta,t)
\int^\theta d\theta'\ A_0(\theta',t) \bigr).
\end{equation}
In terms of the sl(2,R) phase space functions this is
\begin{equation}
q^{(2)}_i
= \int_0^{2\pi} d\theta\ \bigl( - J_i + {1\over 2}C_i^{\ jk}L_j
\int^\theta d\theta'\ L_k \bigr).
\end{equation}
The conservation of this may be checked directly using
(\ref{con}-\ref{cur}):
\begin{eqnarray}
\dot{q}^{(2)}_i &=& \int_0^{2\pi} d\theta\ \bigl[ - L_i'
+ C_i^{\ jk}L_jJ_k
- {1\over 2}C_i^{\ jk}(J_j' \int^\theta d\theta'\ L_k
 + L_j \int^\theta d\theta'\ J_k' )\bigr] \nonumber \\
& = & \int_0^{2\pi} d\theta\ \bigl[C_i^{\ jk}L_jJ_k
+ {1\over 2}C_i^{\ jk}
(J_jL_k - L_j J_k )\bigr] = 0.
\end{eqnarray}
The Poisson bracket of the first two charges is
\begin{equation}
\{q_i^{(1)},q_j^{(2)}\} = C_{ij}^{\ \ k}q_k^{(2)}.
\end{equation}
Since $q^{(1)}_i$ form an sl(2,R)$\sim$so(2,1) Lie algebra it follows
that all the  observables $q^{(n)}_i$ with sl(2,R) indices will
have the Poisson algebra
\begin{equation}
\{q^{(1)}_i, q^{(n)}_j\} = C_{ij}^{\ \ k}q^{(n)}_k
\end{equation}
The Poisson algebra of the higher observables $q^{(n)}_i$ with
themselves
is more involved and there are in general non-linear combinations of
 observables on the right hand sides. We note that given the first
two
 observables $q_i^{(1)},q_j^{(2)}$, the remaining may also be
generated
 by taking Poisson brackets of these with themselves. Another
 feature of this set is that they  are sl(2,R) Lie algebra valued
 whereas the first set obtained above, using $M$ (\ref{mon}), are
traces of
SL(2,R) group elements.

In the steps above, we have obtained a gauge fixed version of complex
two Killing field reduced gravity, and given two methods for
obtaining
observables. The observables are for the complexified theory
and reality conditions must be imposed on them to obtain their
restrictions on the Euclidean or Lorentzian sections.

The restriction to the Euclidean section  involves just setting the
$L_i,J_i$ to be real. The  Lorentzian restriction requires setting
the
triads to be real, and imposing $A_\alpha^I + \bar{A}_\alpha^I =
2\Gamma_\alpha^I(E)$. This reality condition implies that the
 complex conjugate of the observables are also observables. Therefore
 when the triads are set to be real, if $Q[A,E]$ is an observable, so
is
$Q[\bar{A},E]$. Thus $Q[A,E] + Q[\bar{A},E]$ is a {\it real}
observable
for the {\it complex} theory. The real observables for the Lorentzian
section in terms of the original phase space variables may be
obtained as
\begin{equation}
 (Q[A,E] + Q[\bar{A},E])|_{\bar{A} = 2\Gamma - A}.
\end{equation}

 \section{Solution generating technique}

In this section we discuss the relation between the second set of
observables  obtained above and the solution generating technique for
spacetimes with two commuting Killing fields given by Geroch
\cite{geroch}.
We  note only the general features of the method, which are
unchanged by the reality conditions.

 A solution of the Einstein equation with two commuting
 spacelike Killing fields is a phase space trajectory labelled by
values
 of the conserved quantities $ q^{(n)}_i $. A new solution can be
generated
 from a given one by considering the  Hamiltonian flow
 of the phase space variables $L_i,J_i$ generated by the observables
 $ q^{(n)}_i $.  This flow may be parametrized
 by a parameter $s$, and specified by giving three `shift'
 functions $F^i(s)$:
\begin{eqnarray}
{dL_i(t,\theta;s)\over ds} &=& \{ L_i(t,\theta;s), F^k(s) q^{(n)}_k
\}
\nonumber \\
{dJ_i(t,\theta;s)\over ds} &=& \{ J_i(t,\theta;s), F^k(s) q^{(n)}_k
\}.
\label{solgen}
\end{eqnarray}
 Integration of these equations with the  initial condition that
$L_i(t,\theta;s=0),J_i(t,\theta;s=0)$ lie on the given
solution, gives the values of these variables on the new solution
at say, $s=1$.

We therefore see that a new exact solution of the Einstein equations
may be constructed  from a given one by specifying a curve
$\gamma(s)$
($0\le s \le 1$) in a three dimensional vector space with
tangent vector $F^i(s)$, and with $\gamma(0)$ at the origin.
But these are precisely the conditions given by Geroch for
generating new  solutions from a given one \cite{geroch}. In
particular, the intermediate equations (\ref{solgen}) that need to be
integrated as a part of the procedure  are of  exactly
the same form as those present in  ref. \cite{geroch}.
 Thus the infinite number of sl(2,R)   observables obtained in the
last section may be viewed as the phase space analogs of
the generators of Geroch's transformation.

\section{Discussion}

The main new result given in this paper is the explicit construction
of
an infinite number of phase space observables for spacetimes with
 two commuting spacelike Killing vector fields.  The previous studies
 of this reduction of the Einstein equations, in particular Geroch's
 work,  provided strong indications of the existence of such
observables.

Our approach involved rewriting the Hamiltonian evolution equations
using
the Ashtekar variables, and then choosing a particular gauge fixing
which allowed these equations to be rewritten as those of the SL(2,R)
 chiral model (\ref{cons}-\ref{curv}). From this form of the
equations,
 two known methods
were used to obtain the observables. The first made use of the
conservation of the trace of the monodromy matrix $M$ (\ref{mon}),
which acts as the generating functional for the observables. The
  second  made use of a recursive procedure given by
 Brezin et. al. \cite{bizz} to calculate non-local conserved charges
 in two dimensional models.

One set of observables obtained from the monodromy matrix
 have a  structure similar to that of the loop observables
  that have been  used to study the quantization of  full 3+1 gravity
\cite{lc}. This is interesting and suggests that it should be
possible to obtain the quantized two-Killing field reduction directly
from the the full 3+1 observables.

 The second set have an infinite dimensional algebra which doesn't
appear
 to have a simple form. However, as discussed in section VI these
  observables can be used
 to give a solution generating method for this sector of the Einstein
 equations. In particular, the solution generating procedure has
exactly
 the same ingredients as Geroch's one, which indicates that it is the
phase
 space analog of it.

 One of main reasons for addressing the observables problem is that
 it provides one way to address the quantization issue. For generally
 covariant theories  the observables are also the fully gauge
invariant
phase space  variables. A quantum theory may be constructed
by finding a representation of  the Poisson algebra
 of a complete set of classical observables. From the results given
above,
 the second set of observables $q^{(n)}_i$ may be suitable for this
  provided their Poisson algebra can be put into a more  manageable
form.
 Previous work \cite{wu} on a simpler method of obtaining the Geroch
procedure provides a hint that this Poisson algebra may actually be
an
SL(2,R) Kac-Moody (Affine) algebra. The task is then to see if the
$q^{(n)}_i$ can be replaced by some functions of them such that  the
Poisson algebra simplifies to this. This is under investigation.

A further question regarding the observables that hasn't been
addressed
is the question of completeness: Can any invariant phase space
variable
be expressed as a sum of products of the observables obtained here?
In particular, is there any relation between the observables obtained
using
 the two different methods?  These questions are important for
studying
quantization, which has been previously studied in the loop space
representation in ref. \cite{lv}. It was found that there are an
infinite number of  observables in the quantum theory that form a
gl(2) loop
algebra. However, surprisingly the classical counterparts of these
observables was not known. It is likely that the observables given
here
form a subset of these quantum observables, and the correspondence
merits
 further study.

 \acknowledgements
I thank Lee Smolin for an initial collaboration on the two Killing
field reduction discussed here and for many insights. I also
thank Nenad Manojlovic and Charles Torre for discussions. This work
was
supported by the Natural Science and Engineering Research Council of
Canada.

\end{document}